\documentclass[runningheads]{llncs}
\usepackage[T1]{fontenc}
\usepackage{graphicx}
\usepackage{esvect} 
\usepackage{amsmath,amssymb,bm}
\usepackage{algorithm}
\usepackage{algpseudocode}
\usepackage{enumitem}

\usepackage{cite}

\usepackage{url}
\usepackage[colorlinks, bookmarksopen, bookmarksnumbered, citecolor=blue, linkcolor=blue, urlcolor=blue]{hyperref}
\let\subparagraph\relax
\usepackage[capitalize]{cleveref}

\usepackage{titlesec}

\usepackage{etoolbox}
\AtBeginDocument{%
  \setlength{\abovedisplayskip}{9pt plus 2pt minus 4pt}
  \setlength{\belowdisplayskip}{9pt plus 2pt minus 4pt}
}

\begin{document}

\title{A Confidence-Constrained Cloud-Edge Collaborative Framework
for Autism Spectrum Disorder Diagnosis}

\titlerunning{Cloud-Edge Collaborative Framework for ASD Diagnosis}

\author{Qi Deng\inst{1}\orcidID{0009-0000-9024-2246}
\and
Yinghao Zhang\inst{2}\orcidID{0009-0008-2020-869X}
\and
Yalin Liu\inst{1,*}\orcidID{0000-0003-2870-4598}
\and
Bishenghui Tao\inst{1}\orcidID{0000-0003-0968-2346}
}

\institute{Hong Kong Metropolitan University, Hong Kong, China \and
Advanced Institute of Natural Sciences, Beijing Normal University, Zhuhai, China \email{ylliu@hkmu.edu.hk}}

\maketitle            
\setcounter{footnote}{0}

\begin{abstract}
Autism Spectrum Disorder (ASD) diagnosis systems in school environments increasingly relies on IoT-enabled cameras, yet pure cloud processing raises privacy and latency concerns while pure edge inference suffers from limited accuracy. We propose Confidence-Constrained Cloud-Edge Knowledge Distillation (C3EKD), a hierarchical framework that performs most inference at the edge and selectively uploads only low-confidence samples to the cloud. The cloud produces temperature-scaled soft labels and distils them back to edge models via a global loss aggregated across participating schools, improving generalization without centralizing raw data. On two public ASD facial-image datasets, the proposed framework achieves a superior accuracy of 87.4\%, demonstrating its potential for scalable deployment in real-world applications.

\end{abstract}

\section{Introduction}
\vspace{-0.2cm}

Autism Spectrum Disorder (ASD) is a neurodevelopmental condition characterized by early-appearing social communication deficits and repetitive patterns of behavior and interests. ~\cite{fuller2020effects}. Early intervention can significantly improve developmental outcomes for children with ASD, enhancing cognitive abilities, improving daily living skills, reducing symptom severity, and maintaining long-term gains in intellectual ability and language development ~\cite{estes2015long}. In the past few years, ASD diagnosis has emerged as a prominent research focus leveraging technologies including deep learning, multimodal data fusion and Internet of Things (IoT) ~\cite{ahmad2024autism, chawla2023computer,  lakhan2023autism, pavlidis2024federated}.

The deployment of ASD diagnosis systems in school environments has offered significant advantages for early detection and intervention. Schools provide naturalistic observation settings where children's social interactions and communication behaviors can be systematically assessed ~\cite{dekker2016fresh}. School-based screening enables earlier identification, alleviating the burden on specialized clinical facilities that typically impose prolonged waiting time and substantial costs~\cite{gordon2016whittling}. Pan et al. ~\cite{pan2024evaluation} explored AI-driven diagnostic tools for autism detection in school settings, demonstrating the potential for deploying automated screening systems in campus environments.

To effectively deploy ASD diagnosis systems in school environments, several critical requirements need to be addressed. First, low-latency processing is essential for real-time monitoring and timely detection of transient behavioral indicators. Second, diagnostic models need to exhibit robustness across diverse populations, as ASD manifestations vary significantly among different demographic groups and institutions. Third, high diagnostic accuracy must be maintained on resource-constrained edge devices deployed at individual schools, where computational capabilities are inherently limited.

However, existing ASD diagnosis systems face significant challenges when deployed in school environments. First, cloud-based architectures \cite{mohammed2024smart} that process sensor data on remote servers introduce network transmission delays, preventing real-time detection of transient behavioral indicators. Second, models trained on single-institution datasets \cite{atlam2025automated} \cite{reddy2024diagnosis} exhibit limited generalization when deployed across schools with diverse student demographics. Third, lightweight models deployed on resource-constrained edge devices \cite{mahmood2025leveraging} demonstrate relatively lower diagnostic accuracy compared to standard deep learning models.

To address these challenges, we propose Confidence-Constrained Cloud-Edge Knowledge Distillation (C3EKD), a hierarchical framework that fundamentally rethinks ASD diagnosis system for school environments. C3EKD reduces network latency  to enable real-time capture of transient behavior. Furthermore, edge models acquire diagnostic information from  multi-institutional data, ensuring robust generalization across heterogeneous student populations. Moreover, C3EKD maintains high diagnostic accuracy on resource-constrained edge devices.
In summary, the main contributions of this paper are:
\vspace{-0.1cm}
\renewcommand{\labelitemi}{•}
\begin{itemize}

\item We propose C3EKD, a confidence-constraint cloud-edge collaborative framework that performs most diagnostics locally at edge servers and selectively uploads only uncertain samples to the cloud. By avoiding unnecessary cloud communication for high-confidence predictions, this design significantly reduces latency, thus enabling continuous real-time monitoring.

\vspace{0.1cm}
\item We develop a knowledge distillation-based edge model update strategy, where the cloud model progressively transfers diagnostic knowledge to edge models through temperature-scaled soft labels. This enables lightweight edge models to approximate cloud-level accuracy while preserving computational efficiency in resource-constrained environments.
\vspace{0.1cm}
\item We conduct systematic experiments on two public ASD datasets, where the proposed framework achieves 87.4\% accuracy while reducing latency by 7.1\% compared to pure cloud computing. Through 60-round simulation, we demonstrate that the framework achieves near cloud-level accuracy while maintaining a favorable balance between diagnostic accuracy and latency.
\end{itemize}

The remainder of this paper is structured as follows: Section~\ref{sec:system} details the C3EKD framework and its two core mechanisms. Section~\ref{sec:experiment} presents experimental evaluation. Section~\ref{sec:conclusion} concludes the paper.

\section{System Design}
\label{sec:system}
\subsection{C3EKD Framework}
\begin{figure} 
    \centering
    \includegraphics[width=\textwidth]{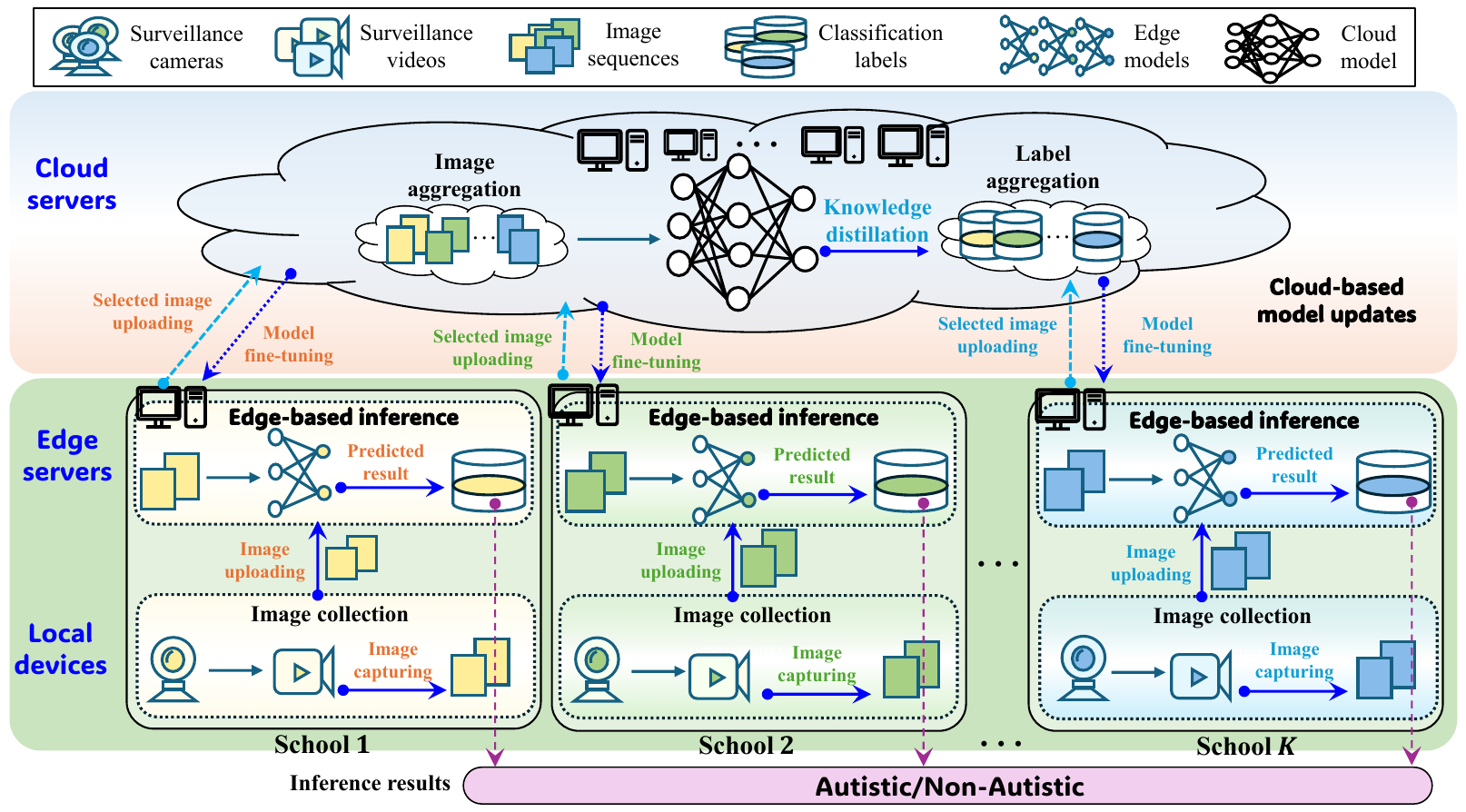}
    \caption{The C3EKD framework for cloud-edge collaborative ASD diagnosis. }
    \label{fig:framework}
\end{figure}

The C3EKD (Confidence-Constrained Cloud-Edge Knowledge Distillation) framework enables efficient and accurate cloud-edge collaborative ASD diagnosis. 
As illustrated in \cref{fig:framework}, our framework adopts a cloud-edge architecture that comprises three distinct infrastructures. 

\vspace{0.1cm}
(i) \textbf{Local devices}: Surveillance cameras deployed at multiple schools continuously capture facial image sequences for analysis.
\vspace{0.1cm}

(ii) \textbf{Edge servers}: Edge servers co-located with individual schools host lightweight neural network models (e.g., MobileNetV2) that perform rapid preliminary inference while preserving data privacy.
\vspace{0.1cm}

(iii) \textbf{Cloud server}: A centralized cloud server equipped with a larger and more complex model (e.g., ResNet-101) that typically achieves higher classification accuracy, providing enhanced inference capabilities and coordinating model updates across the entire system.

The framework performs ASD diagnosis as a binary classification task (ASD vs. non-ASD), i.e., loading the captured surveillance images into a classification model and predicting ASD results. In particular, we design the following two mechanisms to ensure efficient and accurate ASD diagnosis.

\begin{itemize}
    \item \textbf{Mechanism 1 - Edge-based inference}: The edge server is used to execute efficient model inference for fast responses. First, local cameras at each school collect facial image sequences and transmit them to the corresponding edge server. Then the edge server employs a lightweight model to perform local inference, generating preliminary classification results along with confidence scores to evaluate prediction reliability. Because of the lightweight model, the edge inference may have low confidence for some images. 

\vspace{0.2cm}
    \item \textbf{Mechanism 2 - Cloud-based model update}: The cloud server is used to perform parameter updates of edge models for enhanced accuracy. First, the edge server uploads these low-confidence images and their edge soft labels 
    to the cloud server. Then the cloud server, equipped with a larger and more accurate model, aggregates uncertain samples and then performs inference on them to obtain soft labels. The loss between cloud and edge soft labels is distilled back to the edge servers, enabling continuous fine-tuning of edge models and improving their subsequent inference accuracy.

\end{itemize}

\subsection{Mechanism 1 - Edge-based inference}
\label{sub:mech1}
\vspace{-0.1cm}
Consider a deployment with $K$ schools, where school $k$ ($k=1,\dots,K$) operates $J_k$ cameras. The system operates in rounds $r=1,\dots,R$, where each round executes inference and model updating mechanisms. 
Let $I_{k,j,r}$ be the set of images captured by camera $j$ at school $k$ during round $r$. 

Given an image $x\in I_{k,j,r}$, the edge model $f_{\mathrm{edge}}$ first performs local inference to perform ASD diagnosis, producing a binary classification result $y$  as output, where $y \in \{\mathrm{ASD}, \mathrm{Non\text{-}ASD}\}$. To accomplish the binary classification task, we employ the softmax function to generate probability distributions over the two classes, given by
$P_{\mathrm{edge}}(y\mid x) = \mathrm{softmax}\big(f_{\mathrm{edge}}(x)\big)$. The predicted class is determined by selecting the class with maximum probability, given by $\hat{y}_{\mathrm{edge}} = \arg\max_{y} P_{\mathrm{edge}}(y\mid x)$.

To evaluate the certainty of the edge model's prediction for the image $x$, we introduce a confidence metric that quantifies how certain the model is about its prediction. This confidence measure is formulated as:

\begin{equation}
    C(x) = \big|P_{\mathrm{edge}}(\mathrm{ASD}\mid x) - P_{\mathrm{edge}}(\mathrm{Non\text{-}ASD}\mid x)\big|.
    \label{eq:confidence}
\end{equation}
The values of $C(x) \in [0,1]$ represent how confident the edge model is in its prediction, where a larger value indicates a higher certainty. 
The edge server compares $C(x)$ with a threshold $\tau$ to determine whether model updating is needed. Particularly, if $C(x) \geq \tau$, the edge prediction is considered reliable and is accepted without cloud involvement; or if $C(x) < \tau$, the prediction is uncertain, triggering image upload to the cloud for refinement.

\subsection{Mechanism 2 - Cloud-based model update}
\vspace{-0.1cm}
Model update is triggered when edge inference produces low-confidence predictions, requiring cloud assistance for both improved classification and model enhancement. 
The system updates edge models through knowledge distillation\footnote{Knowledge distillation is a model compression technique in which a small model (student) learns from a larger, more accurate model (teacher). Instead of relying solely on true labels, the student model learns from the teacher’s soft label, which convey richer information about class similarities.}. Knowledge distillation enables edge model updates through knowledge transfer from the sophisticated cloud model $f_{\mathrm{cloud}}$ to the lightweight edge model $f_{\mathrm{edge}}$.

To facilitate this knowledge transfer, we employ temperature scaling with a temperature parameter $T$ that smooths the probability distribution to generate more informative soft labels. The soft labels for the low-confidence samples from both edge and cloud models are computed as:

\begin{equation}
    P_{\mathrm{edge}}^{\mathrm{soft}}(y\mid x) = \mathrm{softmax}\!\big(f_{\mathrm{edge}}(x,T)),
    \label{eq:edge_softlabels}
\end{equation}
\begin{equation}
     P_{\mathrm{cloud}}^{\mathrm{soft}}(y\mid x) = \mathrm{softmax}\!\big(f_{\mathrm{cloud}}(x, T)).
     \label{eq:cloud_softlabels}\end{equation}

Based on the produced soft labels, we update the edge model via the knowledge distillation loss function, which combines distillation loss and cross-entropy loss. The distillation loss $L_{\text{KD}}$ measures the KL divergence between cloud and edge soft labels, while the cross-entropy loss $L_{\text{CE}}$ introduces manual clinical annotations  $y_{\mathrm{true}}$ as ground truth labels to guide the model updates. These two losses are combined together to constitute the loss function $L$:

\begin{equation}
    L_{\text{KD}}(x) = \sum_{y} P_{\mathrm{cloud}}^{\mathrm{soft}}(y\mid x) \log {P_{\mathrm{edge}}^{\mathrm{soft}}(y\mid x)},
    \label{eq:kd_loss}
\end{equation}

\vspace{-0.1cm}
\begin{equation}
    L_{\text{CE}}(x) = -\sum_{y} y_{\mathrm{true}} \log P_{\mathrm{edge}}(y\mid x),
    \label{eq:ce_loss}
\end{equation}

\vspace{-0.1cm}
\begin{equation}
    L(x) = \alpha L_{\text{KD}} + (1-\alpha) L_{\text{CE}},
    \label{eq:total_loss}
\end{equation}
where $\alpha$ is a hyper-parameter for balancing the weight of distillation loss and the cross-entropy loss.

To sum up, the cloud aggregates losses across all samples from $K$ schools to compute a global loss, which is given by
\begin{equation}
    L^{r}_{\mathrm{global}} = \frac{1}{K}\sum_{k=1}^{K}\frac{1}{J_k}\sum_{j=1}^{J_k}\frac{1}{|I_{k,j,r}|}\sum_{x\in I_{k,j,r}} L(x).
    \label{eq:global_loss}
\end{equation}
Edge model parameters are then updated via gradient descent:
\begin{equation}
    \Theta_{\mathrm{edge}}^{(r+1)} \leftarrow \Theta_{\mathrm{edge}}^{(r)} - \eta \nabla_{\Theta} L^{r}_{\mathrm{global}},
    \label{eq:update}
\end{equation}
where $\eta$ is the learning rate. The updated parameters are broadcast to all edge servers for the next round. 

\subsection{Complete Workflow}
\cref{alg:ccekl} presents the complete workflow of C3EKD, which integrates the two mechanisms to achieve efficient and accurate cloud-edge collaborative ASD diagnosis. The algorithm operates in rounds to progressively improve edge model performance through iterative inference and model updating.

Each round consists of the execution of two mechanisms. \textbf{Mechanism 1} (lines 5-15) performs edge-based inference with confidence evaluation. High-confidence predictions are accepted locally to ensure low-latency diagnosis, while low-confidence predictions are uploaded to the cloud for refinement. This confidence-constrained strategy minimizes communication overhead while maintaining diagnostic quality. \textbf{Mechanism 2} (lines 16-23) handles cloud-based model updates through temperature-scaled knowledge distillation. Both edge and cloud models employ temperature scaling to generate soft labels that facilitate effective knowledge transfer from cloud model to edge models. For consistent edge-cloud predictions, standalone distillation loss enables direct knowledge transfer, while inconsistent predictions indicating potential misdiagnosis incorporate manual annotations to guide model refinement. This dual-pathway distillation strategy ensures robust model updates while maintaining diagnostic accuracy across diverse ASD manifestations. The aggregated global loss across all schools enables edge models to learn from diverse ASD manifestations, and the updated parameters are then broadcast back to edge servers.

The integration of these two mechanisms addresses the fundamental challenge of achieving high diagnostic accuracy while maintaining real-time responsiveness in distributed school environments. This iterative workflow has three key benefits: (i) it preserves low latency for real-time inference, (ii) it improves edge model generalization capability for heterogeneous populations and (iii) it allows edge models to continuously update and approximate the cloud model accuracy.

\begin{algorithm}[ht]
\caption{Complete workflow of C3EKD}
\label{alg:ccekl}
\begin{algorithmic}[1]
    \State \textbf{Input:} $K,J_k,R,\tau,T,f_{\mathrm{edge}}(\Theta_{\mathrm{edge}})$,$f_{\mathrm{cloud}}(\Theta_{\mathrm{cloud}})$.
    \For{$r = 1,\dots,R$}
        \State $\mathcal{U} \leftarrow \emptyset$ \Comment{set of uncertain images selected for upload}
        \State \textbf{// Mechanism 1: Edge-based inference}
        \For{$k\in \{1,..,K\}$,$j\in \{1,...,J_K\}$}
            \State Collect image sequences $I_{k,j,r}$; transmit $I_{k,j,r}$ to the edge server.
            \For{$x \in I_{k,j,r}$}
                \State Edge compute $P_{\mathrm{edge}}(y\mid x)$, $\hat{y}_{\mathrm{edge}}(x)$ and $C(x)$ (\cref{eq:confidence})
                \If{$C(x) < \tau$}
                    \State add $\big(x, P_{\mathrm{edge}}^{\mathrm{soft}}, \hat{y}_{\mathrm{edge}}(x)\big)$ to $\mathcal{U}$.
                \Else
                    \State Accept $\hat{y}_{\mathrm{edge}}(x)$ locally.
                \EndIf
            \EndFor
        \EndFor
        \State \textbf{// Mechanism 2: Cloud-based model update}
        \State Edge computes $P_{\mathrm{edge}}^{\mathrm{soft}}(y\mid x)$ (~\cref{eq:edge_softlabels}) for all $x \in \mathcal{U}$.
        \State Cloud computes $P_{\mathrm{cloud}}^{\mathrm{soft}}(y\mid x)$  (\cref{eq:cloud_softlabels}) for all $x \in \mathcal{U}$.

        \For{$x \in \mathcal{U}$} 
            \State Request annotation $y_{\mathrm{true}}$, compute loss function $L(x)$ (\cref{eq:kd_loss,eq:ce_loss,eq:total_loss}).

        \EndFor
        \State Compute $L^r_{\mathrm{global}}$ (Eq.~\eqref{eq:global_loss}), update $\Theta_{\mathrm{edge}}^{(r+1)}$ (Eq.~\eqref{eq:update}) to the edge server.
    \EndFor
\end{algorithmic}
\end{algorithm}

\section{Experiments and Results}
\label{sec:experiment}

\vspace{-0.2cm}
\subsection{Experimental Setup}

\subsubsection{Dataset.}
Experiments are conducted on a combination of two publicly datasets from Kaggle. The first dataset \footnote{\url{https://www.kaggle.com/datasets/imrankhan77/autistic-children-facial-data-set?select=consolidated}}contains 2,936 images, while the second dataset\footnote{\url{https://www.kaggle.com/datasets/shafi420/facial}} includes 2,940 images. Each image is labeled as either ``\textit{Autistic}'' or ``\textit{Non-Autistic}''. All images are resized to $224 \times 224$ pixels. Here, each image size is set to 30kB and the bandwidths of local devices-to-edge server and edge server-to-cloud server are set to 5 Mbps and 20 Mbps, respectively.

\vspace{-0.2cm}
\subsubsection{Implementation Details.}
We partition the dataset into training set (2,352 images, 40\%), simulation set (2,880 images, 50\%), and testing set (644 images, 10\%). Initial edge and cloud model training are performed on the training set independently, establishing baseline performance with edge models (MobileNet-v2) achieving 71.95\% accuracy and cloud model (ResNet-101) achieving 89.46\% accuracy. We simulate a scenario with three edges, where each edge is connected to one camera. Each camera would capture 16 images and then upload them to the edge server 
at the beginning of each round. Within each round, we record communication overhead associated with our edge-based inference (Mechanism 1) and cloud-based model update (Mechanism 2) in the simulation set and evaluate the model inference accuracy by comparing the predicted labels produced by our framework against the ground truth labels for each sample in the testing set. We conduct 60 simulation rounds to emulate the continuous operation of our framework, where edge servers persistently collect and process facial images while adaptively updating their models through cloud collaboration. Please refer to our GitHub\footnote{\url{https://github.com/Qi-Deng-HKMU/C3EKD_Experiments}} for our source code and experiment data.

\vspace{-0.3cm}
\subsubsection{Evaluation Metrics.}
To examine the performance of our proposed C3EKD framework, we define the following metrics:

\vspace{-0.2cm}
\begin{itemize}
    \item \textbf{Accuracy.} The ratio of correctly classified samples to total samples.

    \item \textbf{Relative Accuracy.} The quotient of the edge model’s standalone accuracy and the framework’s overall accuracy

    \item \textbf{Upload Proportion.} The ratio of samples requiring upload to the cloud server to the total number of samples across all 60 simulation rounds, reflecting communication efficiency.

    \item \textbf{Average Time Delay.} It is defined as:  $\bar{T} = \frac{1}{N} \sum_{i=1}^{N} T_{\text{trans}}$, where $T_{\text{trans}}$ is the transmission delay, and $N$ is the total number of processed images. 

\end{itemize}

\vspace{-0.3cm}
\subsection{Baselines}
\vspace{-0.1cm}
We compare our proposed method with three computing paradigms:
\vspace{-0.1cm}
\begin{itemize}
    \item \textbf{Pure Edge.} All samples are processed only by edge models; no uploads or updates. This paradigm isolates edge inference 
    to validate the effectiveness of cloud edge collaborative inference. 
    \item \textbf{Pure Cloud.} All samples are uploaded for cloud inference; edge models are unused. This paradigm bypasses local edge inference entirely to validate the effectiveness of the edge-based inference (Mechanism 1).
    \item \textbf{Collaborative (w/o Update).} Edge first infers. If $C(x)\!\ge\!\tau$, edge prediction is final; otherwise, upload to cloud for inference. No edge update between rounds. This paradigm validates the effectiveness of cloud-based model updates (Mechanism 2) by isolating edge model update strategy.

\end{itemize}

\begin{table}[t!]
\centering
\caption{Computing Paradigms Performance Comparison}
\label{tab:paradigm_comparison}
\begin{tabular}{|l|c|c|c|}
\hline
Computing Paradigms& Upload Proportion & Accuracy & Time Delay (ms) \\
\hline
Pure Edge & 0.00 & 68.5\% & \textbf{12.29} \\
Pure Cloud & 1.00 & \textbf{87.8\%} & \textbf{14.75} \\
Collaborative Framework ($\tau$=0.1) & 0.324 & 76.7\% & 13.08 \\
Collaborative Framework ($\tau$=0.2) & 0.618 & \textbf{80.1}\% & 13.81 \\
Collaborative Framework ($\tau$=0.3) & 0.825 & 81.8\% & 14.31 \\
Proposed C3EKD ($\tau$=0.1) & 0.239 & 79.8\% & 12.88 \\
Proposed C3EKD ($\tau$=0.2) & 0.416 & \textbf{85.7}\% & 13.31 \\
Proposed C3EKD ($\tau$=0.3) & 0.577 & \textbf{87.4\%} & \textbf{13.71} \\

\hline
\end{tabular}
\end{table}

\vspace{-0.2cm}
\subsection{Results}
\subsubsection{Overall Evaluation.} 
\cref{tab:paradigm_comparison} summarizes the performance across different paradigms. Pure cloud computing delivers the highest accuracy (87.8\%) but incurs the largest upload cost, as every sample is transmitted. Pure edge computing offers the lowest latency (12.29 ms) but sacrifices accuracy, achieving only 68.5\%. The collaborative framework without model updates improves accuracy to 80.1\% at $\tau{=}0.2$, confirming that selective upload alleviates edge model limitations, but its performance soon saturates. By contrast, our proposed C3EKD consistently outperforms this baseline: with $\tau{=}0.2$, accuracy rises to 85.7\% while maintaining similar delay, directly evidencing the benefit of cloud-based updates. Furthermore, C3EKD at $\tau{=}0.3$ nearly matches pure cloud accuracy (87.4\% vs. 87.8\%) while reducing latency by 7.1\%, showing that the framework achieves a favorable balance between accuracy and efficiency. These results demonstrate that the introduction of iterative updates is the critical factor enabling edge models to approximate cloud-level accuracy without excessive communication.

\vspace{-0.2cm}
\subsubsection{Convergence Analysis.} 
\begin{figure}
    \centering
    \includegraphics[width=\textwidth]{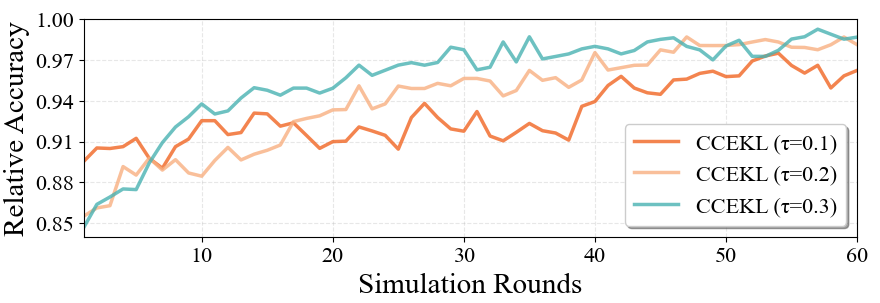}
    \caption{Relative accuracy (rAcc) progression over 60 communication rounds demonstrating edge model convergence toward cloud-level performance.}
    \label{fig:convergence_analysis}
\end{figure}

\cref{fig:convergence_analysis} illustrates the progression of relative accuracy (rAcc) in 60 rounds under different thresholds. All settings show clear upward trends, with rAcc gradually approaching 100\%, confirming that that edge models progressively align with the cloud-level accuracy. 
The higher threshold accelerates this process: $\tau{=}0.3$ achieves the best convergence because more uncertain samples are uploaded, allowing edge models to learn from richer supervision. Occasional fluctuations appear, which are expected in distributed learning with stochastic sampling, but the long-term trajectory remains stable. Overall, these results validate that C3EKD not only improves snapshot performance but also ensures steady convergence of edge models toward cloud-level accuracy across communication rounds.

\vspace{-0.2cm}

\vspace{-1.0cm}
\section{Conclusion}
\label{sec:conclusion}

\vspace{-0.2cm}
In this paper, we propose C3EKD, a novel cloud-edge collaborative framework for edge-based facial image classification, which achieves efficient edge model update by leveraging both edge inference and cloud-assisted knowledge distillation. C3EKD selectively uploads only uncertain edge predictions to the cloud, enabling the cloud server to aggregate all samples from distributed edge servers. To address the challenge of edge model performance under limited local computation, the cloud model generates temperature-scaled soft labels to guide edge learning, allowing continuous enhancement of edge model accuracy while minimizing time delay. Moreover, the edge model update mechanism adopts knowledge distillation, effectively addressing prediction inconsistencies and enhancing model generalization across heterogeneous edge devices. Experiments on ASD image classification have demonstrated the effectiveness of the proposed framework in both classification performance and communication efficiency. 
C3EKD provides a promising approach for privacy-preserving edge AI in medical imaging and can be applied to various real-time healthcare and diagnostic applications without data leakage. 

\vspace{-0.2cm}
\subsubsection{Acknowledgements.} The work in this paper was supported in part by the grant from the Research Grants Council of the Hong Kong Special Administrative Region, China, under project No. UGC/FDS16/E15/24, and in part by the UGC Research Matching Grant Scheme under Project No.: 2024/3003.

\bibliographystyle{splncs04}
\bibliography{ref}

\begin{thebibliography}{10}
\providecommand{\url}[1]{\texttt{#1}}
\providecommand{\urlprefix}{URL }
\providecommand{\doi}[1]{https://doi.org/#1}

\bibitem{ahmad2024autism}
Ahmad, I., Rashid, J., Faheem, M., Akram, A., Khan, N.A., Amin, R.u.: Autism spectrum disorder detection using facial images: A performance comparison of pretrained convolutional neural networks. Healthcare Technology Letters  \textbf{11}(4),  227--239 (2024)

\bibitem{atlam2025automated}
Atlam, E.S., Aljuhani, K.O., Gad, I., Abdelrahim, E.M., Atwa, A.E.M., Ahmed, A.: Automated identification of autism spectrum disorder from facial images using explainable deep learning models. Scientific Reports  \textbf{15}(1),  26682 (2025)

\bibitem{chawla2023computer}
Chawla, P., Rana, S.B., Kaur, H., Singh, K.: Computer-aided diagnosis of autism spectrum disorder from eeg signals using deep learning with fawt and multiscale permutation entropy features. Proceedings of the Institution of Mechanical Engineers, Part H: Journal of Engineering in Medicine  \textbf{237}(2),  282--294 (2023)

\bibitem{dekker2016fresh}
Dekker, V., Nauta, M.H., Mulder, E.J., Sytema, S., de~Bildt, A.: A fresh pair of eyes: A blind observation method for evaluating social skills of children with asd in a naturalistic peer situation in school. Journal of Autism and Developmental Disorders  \textbf{46}(9),  2890--2904 (2016)

\bibitem{estes2015long}
Estes, A., Munson, J., Rogers, S.J., Greenson, J., Winter, J., Dawson, G.: Long-term outcomes of early intervention in 6-year-old children with autism spectrum disorder. Journal of the American Academy of Child \& Adolescent Psychiatry  \textbf{54}(7),  580--587 (2015)

\bibitem{fuller2020effects}
Fuller, E.A., Kaiser, A.P.: The effects of early intervention on social communication outcomes for children with autism spectrum disorder: A meta-analysis. Journal of autism and developmental disorders  \textbf{50}(5),  1683--1700 (2020)

\bibitem{gordon2016whittling}
Gordon-Lipkin, E., Foster, J., Peacock, G.: Whittling down the wait time: exploring models to minimize the delay from initial concern to diagnosis and treatment of autism spectrum disorder. Pediatric Clinics of North America  \textbf{63}(5), ~851 (2016)

\bibitem{lakhan2023autism}
Lakhan, A., Mohammed, M.A., Abdulkareem, K.H., Hamouda, H., Alyahya, S.: Autism spectrum disorder detection framework for children based on federated learning integrated cnn-lstm. Computers in Biology and Medicine  \textbf{166},  107539 (2023)

\bibitem{mahmood2025leveraging}
Mahmood, M.A., Jamel, L., Alturki, N., Tawfeek, M.A.: Leveraging artificial intelligence for diagnosis of children autism through facial expressions. Scientific Reports  \textbf{15}(1),  11945 (2025)

\bibitem{mohammed2024smart}
Mohammed, M.A., Alyahya, S., Mukhlif, A.A., Abdulkareem, K.H., Hamouda, H., Lakhan, A.: Smart autism spectrum disorder learning system based on remote edge healthcare clinics and internet of medical things. Sensors (Basel, Switzerland)  \textbf{24}(23), ~7488 (2024)

\bibitem{pan2024evaluation}
Pan, Y., Foroughi, A.: Evaluation of ai tools for healthcare networks at the cloud-edge interaction to diagnose autism in educational environments. Journal of Cloud Computing  \textbf{13}(1), ~39 (2024)

\bibitem{pavlidis2024federated}
Pavlidis, N., Perifanis, V., Briola, E., Nikolaidis, C.C., Katsiri, E., Efraimidis, P.S., Filippidou, D.E.: Federated anomaly detection for early-stage diagnosis of autism spectrum disorders using serious game data. arXiv preprint arXiv:2410.20003  (2024)

\bibitem{reddy2024diagnosis}
Reddy, P.: Diagnosis of autism in children using deep learning techniques by analyzing facial features. Engineering Proceedings  \textbf{59}(1), ~198 (2024)

\end{thebibliography}

%
%
%
%

\end{document}